\documentclass{aa}  
\usepackage{graphicx}
\usepackage[version=3]{mhchem}
\usepackage{txfonts}
\usepackage{gensymb}
\usepackage{hyperref}
\usepackage{float}
\usepackage[version=3]{mhchem}
\usepackage{threeparttable}
\usepackage{xcolor}
\newlength{\dzero}
\newcommand{\minnumb}{\settowidth{\dzero}{$-$}\kern-\dzero$-$}
\newcommand{\spcn}[1]{\settowidth{\dzero}{0}\kern#1\dzero}
\newcommand{\dgr}{$^{\mbox{$\circ$}}$}

\def\aap{A\&A}
\def\apjs{ApJS}
\def\dda{CHD$_{2}$CHO}

\hypersetup{
    colorlinks=true,
    linkcolor=blue,
    filecolor=magenta,
    citecolor=blue,
    urlcolor=blue,
    pdftitle={Overleaf Example},
    pdfpagemode=FullScreen,
    }
\begin{document} 
\title{Millimetre and sub-millimetre spectroscopy of doubly deuterated\vspace{1.0ex}\\acetaldehyde (\ce{CHD2CHO})
and first detection\vspace{1.0ex}\\towards IRAS 16293-2422}
\titlerunning{Millimetre and sub-millimetre spectroscopy of doubly deuterated acetaldehyde (\ce{CHD2CHO})}

\author{J.~Ferrer Asensio\inst{1}, S.~Spezzano\inst{1}, L.~H.~ Coudert\inst{2}, V.~Lattanzi\inst{1}, C.~P.~Endres\inst{1}, J.~K.~Jørgensen\inst{3}, P.~Caselli\inst{1}}
\institute{\tiny{\inst{1} Center for Astrochemical Studies, Max-Planck-Institut f\"ur extraterrestrische Physik, Giessenbachstr. 1, 85748 Garching, Germany\\  \inst{2}Institut des Sciences Mol\'eculaires d’Orsay (ISMO), CNRS, Universit\'e Paris-Sud, Universit\'e Paris-Saclay, F-91405 Orsay, France \\ \inst{3} Niels Bohr Institute, University of Copenhagen, {\O}ster Voldgade 5--7, DK-1350 Copenhagen K., Denmark}}

\authorrunning{J. Ferrer Asensio et al.}

\date{Received  ; accepted }

\abstract
%context
{The abundances of deuterated molecules with respect to their main isotopologue counterparts have been determined to be orders of magnitude higher than expected from the cosmic abundance of deuterium relative to hydrogen. The increasing number of singly and multi-deuterated species detections helps us to constrain the interplay between gas-phase and solid-state chemistry and to understand better deuterium fractionation in the early stages of star formation. Acetaldehyde is one of the most abundant complex organic molecules (COMs) in star-forming regions and its singly deuterated isotopologues have already been observed towards protostars.}
%aims
{A spectroscopic catalogue for astrophysical purposes is built for doubly deuterated acetaldehyde (\dda) from measurements in the laboratory. With this accurate catalogue we aim to search and detect this species in the interstellar medium and retrieve its column density and abundance.}
%approach
{Submillimetre wave transitions were measured for the non-rigid doubly deuterated acetaldehyde \ce{CHD2CHO} displaying hindered
internal rotation of its asymmetrical \ce{CHD2} methyl group. An analysis of a dataset consisting of previously measured microwave transitions and of the newly measured ones was carried out with an effective Hamiltonian which accounts for the tunneling of the asymmetrical methyl group.}
%results
{A line position analysis is carried out allowing us to reproduce 853 transition frequencies with a weighted root mean square standard deviation of 1.7, varying 40 spectroscopic constants. A spectroscopic catalogue for astrophysical purposes is built from the analysis results. Using this catalogue we were able to detect for the first time \ce{CHD2CHO} towards the low-mass protostellar system IRAS 16293-2422 utilizing data from the ALMA Protostellar Interferometric Line Survey.}
%conclusions
{The first detection of the \ce{CHD2CHO} species allows for the derivation of its column density with a value of 1.3$\times$10$^{15}$ cm$^{-2}$ and an uncertainty of 10-20\%. The resulting D$_{2}$/D ratio of $\sim$ 20\% is found to be coincident with D$_{2}$/D ratios derived for other complex organic molecules towards IRAS~16293-2422, pointing at a common formation environment with enhanced deuterium fractionation.}

\keywords{ISM: molecules - ISM: clouds - ISM: abundances - radio lines: ISM - stars: formation - radiative transfer - line: identification}

\maketitle

\section{Introduction} \label{introduction}

The number of multi-deuterated molecules detected in the ISM has increased substantially in the last years (e.g. CHD$_{2}$OH and CD$_{3}$OH \citep{parise:02, parise:04}, c-C$_{3}$D$_{2}$ \citep{spezzano:13}, \ce{D2CO} \citep{turner:90}, \ce{CHD2OHCHO} \citep{manigand:19}, \ce{D2O} \citep{butner:07}, \ce{CH3OCHD2} \citep{richard:21}). In most recent years, the Atacama Large Millimeter/submillimeter Array (ALMA) telescope opened up the possibility of measuring the abundances of these species in a more unambiguous manner. The high spatial-resolution of the interferometre allowed zooming-in into the warm gas around protostars where these molecules appear. The ALMA Protostellar Interferometric Line Survey (PILS) towards the proto-stellar system IRAS 16293-2422 \citep{jorgensen:16} has allowed for the detection and accurate column density derivation of several multi-deuterated species (e.g. \citealp{persson:18},  \citealp{jensen:21}). Doubly-deuterated molecules are found to be more abundant than expected when taking into account the local ISM deuterium abundance (D/H = 2.0 ± 0.1 $\times$ 10$^{-5}$, \cite{linsky:03}; \cite{caselli:12b}; \cite{ceccarelli:14} and references therein). The enrichment of molecules with deuterium, known as deuterium fractionation, is an interesting diagnostics tool that can be used as a clock to trace molecules to the time and environment of their formation \citep{ceccarelli:14}. For example, the D/H ratio for methanol found in comets agrees with the ratio derived in pre-stellar cores and low-mass protostellar regions linking the cometary methanol to the first stages of star formation \citep{drozdovskaya:21}. Furthermore, the water D/H ratio on Earth is found to be more similar to the one observed in proto-stellar cores, in clustered star-forming regions, than that in isolated proto-stellar cores \citep{jensen:19}, supporting the interpretation that the Sun was formed in a clustered star-forming environment \citep{adams:10}. \ 

Deuteration is most effective in the pre-stellar core environment due to the low temperatures present ($<10$ K) (e.g. \citealp{caselli:02b, crapsi:07}). Due to the lower zero point energy (ZPE) of deuterium, this forms stronger bonds than hydrogen at low temperatures, making the equilibrium of the reaction \ce{H3+ + HD <=> H2D+ + H2} to be shifted to the right-hand side in pre-stellar cores (e.g. \citealp{pagani:92}). Moreover, in these environments, CO which is the main destructor of H$_{3}^{+}$ and \ce{H2D+}, is heavily frozen onto the surface of dust grains (e.g. \citealp{caselli:99}). As a consequence, pre-stellar cores have a higher \ce{H2D+ / H3+} ratio \citep{dalgarno:84}. \ce{H2D+} and other multideuterated forms of \ce{H3+} are the main deuteration agent in the gas phase. On the other hand, H-D substitution reactions on the surface of dust grains have been proposed to explain the observed deuterium fractionation (\citealp{drozdovskaya:22} and references therein). In \cite{ambrose:21} deuterated methanol (CH$_{2}$DOH) was observed towards nine of the 12 starless and pre-stellar sources observed, deriving a median value [CH$_{2}$DOH]/[CH$_{3}$OH] ratio of 0.11.   \ 

Deuterated molecules found in sources at later stages of the star formation process are thought to be inherited from the pre-stellar core phase. The molecules trapped in the ice mantles of dust grains are released into the gas phase due to the heating of the central protostar, making their detection possible \citep{taquet:14}. Observation of the D/H ratios towards the proto-stellar system IRAS 16293-2422 have revealed a generalised trend with smaller molecules (e.g. methanol, formic acid, formaldehyde) having a D/H $\sim$2\% and larger molecules (e.g. dimethyl ether, ethanol) displaying a D/H $\sim$4--8\% \citep{jorgensen:18}. \cite{vangelder:22} compared observations of deuterated methanol towards high-mass protostars with literature observations encompassing multiple stages and masses of the star formation process. They observe that the [CH$_{2}$DOH]/[CH$_{3}$OH] ratio for high-mass protostars is lower than the one for the low-mass protostars. However, the [CHD$_{2}$OH]/[CH$_{2}$DOH] ratio is found to be similar amongst high- and low-mass protostars. In the same paper, by using the gas-grain chemical model GRAINOBLE \citep{taquet:12, taquet:13, taquet:14}, the authors suggest that the methanol deuteration levels in high-mass protostars could indicate that this molecule was formed in a warm environment (>20K) or that the pre-stellar phase within which they formed was short lived.\ 

The fact that D$_{2}$/D ratios are observed to be higher than D/H ratios implies that multiple deuteration is more favourable than the first deuteration, which is supported by laboratory experiments \citep{nagaoka:05, nagaoka:07, hidaka:09}. In the case of methanol, from the observed D/H and D$_{2}$/D column density ratios in the comet 67P/Churyumov–Gerasimenko, the formation of singly deuterated methanol (CH$_{2}$DOH) is constrained to happen via the H-D substitution of the main isotopologue (CH$_{3}$OH) \citep{drozdovskaya:21}. On the other hand, doubly deuterated methanol (CHD$_{2}$OH) is deduced to form from the hydrogenation of doubly deuterated formaldehyde (D$_{2}$CO) \citep{drozdovskaya:21}. The study of methanol deuteration sets an example on the importance of deriving column density ratios of singly and multi-deuterated species for the purpose of understanding the nature of deuterium fractionation, and the interplay of the chemistry in the gas phase and on the surface of dust grains.\ 

The astrophysically relevant molecule acetaldehyde and its isotopologues have been the focus of numerous spectroscopic studies due to the internal rotation of their methyl group. The microwave spectrum of the main isotopologue (CH$_{3}$CHO) was first analysed by \citet{kilb:57}. Subsequently, its analysis was extended up to the $\upsilon_t=4$ torsional state \citep{hershbach:59, iijima:72, bauder:76} enabling its first detection in the interstellar medium (ISM) by \citet{gilmore:76}. The isotopic species with either a deuterated \ce{COD} aldehyde group or a fully deuterated \ce{CD3} methyl group were also investigated \citep{coudert:06, elkeurti:10, zaleski:17} of which the \ce{CH3CDO} species has been detected in the ISM by \cite{jorgensen:18}. Spectroscopic results are also available for isotopic species with a partially deuterated \ce{CH2D} or \ce{CHD2} asymmetrical methyl group.  The monodeuterated species \ce{CH2DCHO} has been the subject of several investigations \citep{turner:76, turner:81, coudert:19} which led to its detection in the ISM \citep{coudert:19}. The doubly-deuterated species \ce{CHD2CHO} has also been studied \citep{turner:76, turner:81}, but only a few transitions characterised by low $K_a$-values were assigned in a microwave spectrum. Due to the high levels of confusion in the observational spectra towards star-forming regions, high-accuracy spectroscopic catalogues are crucial for the detection of species in the ISM, which stresses the need to extend the study on \dda \ beyond the work done by \cite{turner:76, turner:81} towards higher J and \ce{K_a}. \ 

The doubly deuterated isotopic variant of acetaldehyde \ce{CHD2CHO} is investigated in this article. The analysis of its microwave and submillimetre wave spectra is reported in Section~\ref{spctr_cd2hcoh}, where a spectroscopic catalogue is also built. Section~\ref{obs} deals with the astrophysical search in the ALMA Protostellar Interferometric Line Survey (PILS) and the detection of this species. In Section~\ref{discussion} we present the discussion.  Lastly, our conclusions can be found on Section \ref{concl}.\

\section{\label{spctr_cd2hcoh}Spectroscopic investigation of \ce{CHD2CHO}}
Theoretical models aimed at accounting for the internal rotation
of molecules displaying internal rotation of a
symmetrical \ce{CH3} or \ce{CD3} methyl group were developed a long time ago \citep{koehler:40, burkhard:51, ivash:53, hecht:57, hecht:57b, lees:68, delucia:89} and successfully applied to the main isotopic species of methanol and acetaldehyde. The efforts to characterise the internal rotation of symmetrical methyl groups is still under study \citep{ilyushin:20, kleiner:20, xu:21}. These models cannot be used for molecules displaying internal rotation of a partially deuterated \ce{CH2D} or \ce{CHD2} methyl group. Alternate models were designed for such molecules and applied to mono- and doubly-deuterated methyl formate and methanol \citep{margules:09, coudert:12, pearson:12, coudert:14, ndao:15, coudert:21}. The Hamiltonian used in the present investigation is based on the theoretical model developed for monodeuterated methyl formate by \citet{margules:09} which relies on the high-barrier internal axis method (IAM) approach of \citet{hougen:85} and \citet{coudert:88}.

In this section, the experimental spectrum is described
and, after briefly outlining the IAM approach, the fitting of previously available microwave transitions \citep{turner:76, turner:81} and of the newly measured submillimetre ones is reported.

\subsection{\label{exp}Experimental}
The rotational transitions were recorded in the 82.5--450~GHz frequency range using the broadband Chirped-Pulse Fourier Transform Spectrometre (CP-FTS) as well as the high-resolution absorption experiment in the Center for Astrochemical Studies Laboratory of the Max-Planck-Institute für Extraterrestrische Physik in Garching, Germany. 

The doubly deuterated acetaldehyde sample was synthesised by warming up a mixture of equal weights of CH$_{3}$CHO and D$_{2}$O in acidic medium (KHSO$_{4}$, pH=1) with a silicone bath at 38\degree C for 2 weeks. Separation of the organic phase, where the molecule is dissolved in, from the water phase was done by manual decantation. The first low $J$ and $K_a$ line recordings were done with the CP-FTS that allows for an instantaneous bandwidth of 20 GHz in the frequency range of 75–110 GHz. The Chirped-Pulse is produced by an arbitrary waveform generator (Keysight, M8190A). The signal is then upconverted and amplified by an IQ modulator and a solid state amplifier respectively before entering the chamber. Thereafter, the signal is amplified,  downconverted and digitised.\

\begin{figure}
\centering
\includegraphics[width=0.9\linewidth]{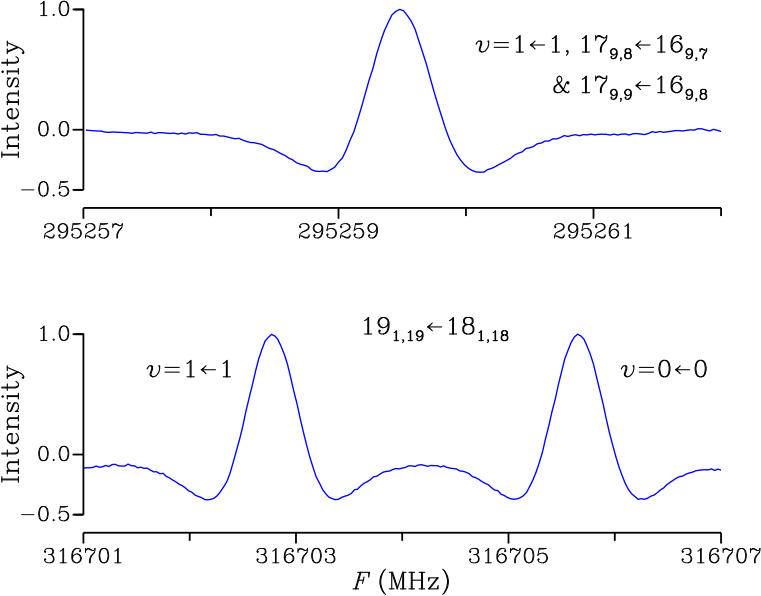}
\caption{\label{mw_sub} The $\upsilon$ = 0 and 1 tunneling components, arising from the two isoenergetic Out configurations, of several $a$-type transitions. The upper panel shows the $\upsilon$ = 1 component of the unresolved K-type doublet $17_{9} \leftarrow 16_{9}$. The lower panel depicts the $\upsilon$ = 0 and 1 tunneling components of the $19_{1,19} \leftarrow  18_{1,18}$ transition. }
\end{figure}

\begin{figure}
\centering
\includegraphics[width=10cm]{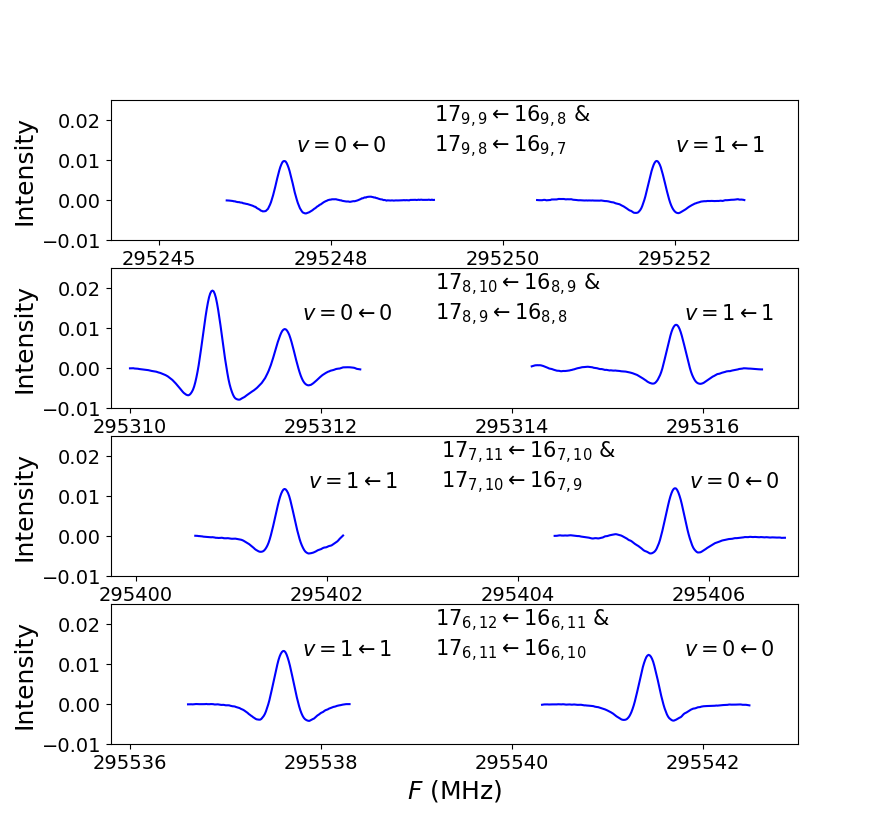}
\caption{\label{branch} The $\upsilon$ = 0 and 1 tunneling components of the $a$-type $17{K_a} \leftarrow  16{K_a}$ transitions, displaying no resolved asymmetry splitting, with 6 $\leq K_{a} \leq$ 9. The line at 295 312 MHz in the second panel is unidentified.}
\end{figure}

For lines at higher $J$ and $K_a$, which require an increased sensitivity, we moved on to recording with the frequency modulated absorption spectrometre \citep{bizzocchi:17}. The radiation source is an active multiplier chain (Virginia Diodes, Inc.) connected to a synthesiser (Keysight E8257D PSG Analog Signal Generator) operating between 250 kHz and 67 GHz. The synthesiser is also connected to a 10 MHz rubidium frequency clock. A combination of frequency multipliers allows us to access the range between 82.5 and 450 GHz covered by the measurements. The detector used is a liquid-He-cooled InSb hot electron bolometre (QMC Instr. Ltd.). Frequency modulation of the signal is applied to reduce the noise, and then the output signal is demodulated at 2$f$ (where $f$ denotes the modulation frequency) with a lock-in amplifier (Standford Research Systems SR830). The sample was at an average pressure of $1.2\times10^{-2}$ mbar in the cell during measurements with both setups and the linewidth was limited by Doppler broadening. All the measurements were carried out at room temperature. Figure~\ref{mw_sub} shows a sample of the measurement scans and  Figure~\ref{branch} shows the Out configuration $a$-type $17_{K_a} \rightarrow 16_{K_a} $ transitions for $K_{a}$ ranging from 6 to 9 (see Figure \ref{confs} for a structural reference of the In and Out configurations.)

\subsection{\label{theo}Theory}
\begin{figure}
\centering
\includegraphics[width=0.9\linewidth]{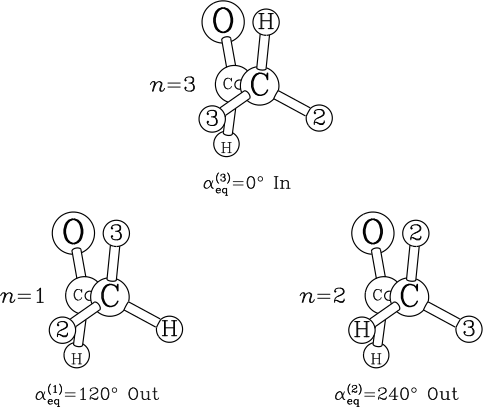}
\caption{\label{confs}The two energetically equivalent Out
configurations and the higher energy In configuration are
identified by their configuration number $n=1,$ 2, and 3.
The two deuterium atoms are labeled 2 and 3.  $\alpha_{\rm
eq}^{(n)}$ is the equilibrium value of the torsional angle
$\alpha=\protect\angle\ce{HCCO}$.  Configuration~3, displaying a symmetry plane and therefore having $C_s$ symmetry, is approximately 14.487~cm$^{-1}$ above Configurations~1 and 2 having $C_1$ symmetry.}\end{figure}

The theoretical model developed previously for monodeuterated methyl formate \citep{margules:09} can be applied to doubly deuterated acetaldehyde \ce{CHD2CHO} with only a few changes.  The main one concerns the relative energy of the non-superimposable equilibrium configurations, defined in agreement with the IAM approach of \citet{hougen:85} and \citet{coudert:88}.  As emphasised by Fig.~\ref{confs}, in doubly deuterated acetaldehyde, just like in monodeuterated methyl formate, there arise three equilibrium configurations which can be identified by their configuration number $n$, with $n=1$, 2, and 3, and characterised by $\alpha_{\rm eq}^{(n)}$ the value of the torsional angle $\alpha=\angle\ce{HCCO}$ about which the reference function is localised.  Configurations~1 and 2 are the two $C_1$ symmetry Out configurations with the hydrogen atom outside the \ce{CCO} plane.  They are lower in energy than Configuration~3, the $C_s$ symmetry In configuration with the hydrogen atom in the \ce{CCO} plane. The energy difference $E_d$ between the In and Out configurations is not known exactly yet but is expected to be very close to the zero-point vibrational energy difference:
\begin{equation}
E_d=E_{\rm zpe}({\rm In,}\, \ce{CHD2CHO}) - E_{\rm zpe}({\rm Out,}\, \ce{CHD2CHO})\;.
\end{equation}
An approximate value of this difference was retrieved from $E_{d}'$, the equivalent energy difference for the monodeuterated species \ce{CH2DCHO} (we refer to Figure 2 in \citealp{coudert:19} for a visual representation of these conformers):
\begin{equation}
E_{d}'={E_{\rm zpe}({\rm Out,}\, \ce{CH2DCHO}) - E_{\rm zpe}({\rm In,}\, \ce{CH2DCHO})}\;.
\end{equation}
The ratio $r=E_d/E_{d}'$ was computed using {\em ab
initio} calculations.  A calculation at the B3LYP/6-31G(d)
level of theory with the Gaussian~16 package \citep{frisch:16} yielded $r=0.9311$.  Since $E_{d}'$, first estimated by \citet{turner:81} and \citet{cox:03} and determined later with a higher accuracy by \citet{coudert:19}, is $15.558\,66(4)$~cm$^{-1}$, we obtain $E_d= 14.487$~cm$^{-1}$.

The theoretical results in Sections~3.2 and 3.3 of \citet{margules:09} can be used in the case of \ce{CHD2CHO} provided a few changes, due to the definition of $E_d$ in this work, are made. Equation~(8) of these authors should be changed into:
\begin{equation}
\langle\psi_3|H_t|\psi_3\rangle= 
\langle\psi_1|H_t|\psi_1\rangle + E_d= 
\langle\psi_2|H_t|\psi_2\rangle + E_d
\end{equation}
and $E_d$ should be ignored in their Eq.~(21) and in their Table~2; in their
Table~1, it should only appear for diagonal matrix elements involing two wavefunctions corresponding to Configuration~3. Table~\ref{rotdep} of the present paper lists computed values for the rotational constants and dipole moment components of the In and Out configurations as retrieved from the structure of \citet{kilb:57} and the dipole moment components reported in Table~16 of \citet{turner:78} for \ce{CH3CHO}. Equations~(12) and (13) of \citet{margules:09} should be used with no change to obtain the tunneling matrix element $H_{JK\gamma 1;JK'\gamma'2}$ of the $1\rightarrow 2$ tunneling path
connecting the isoenergetic Configurations~1 and 2. Similarly, Eqs.~(14) and (15) should be used for tunneling matrix element $H_{JK\gamma 1;JK'\gamma'3}$ of the $1\rightarrow 3$ tunneling path connecting Configurations~1 and 3.  The rotational dependence of these tunneling matrix elements is parameterised by two sets of Eulerian-type angles, $\theta_2,\phi_2$ and $\chi_3,\theta_3,\phi_3$, which were numerically evaluated using the structure of \citet{kilb:57} and which are also given in
Table~\ref{rotdep}.  In Eqs.~(12)--(15) of \citet{margules:09}, $h_2$ and $h_3$ are the magnitude of the tunneling splittings. These parameters, the Eulerian-type angles $\theta_2, \phi_2, \chi_3, \theta_3, \phi_3$, the rotational constants of the In and Out
configurations, and the energy difference $E_d$ allows us to compute to zeroth-order the rotation-torsion energy of the three first torsional states of \ce{CHD2CHO}.

\begin{figure}
\centering
\includegraphics[width=0.9\linewidth]{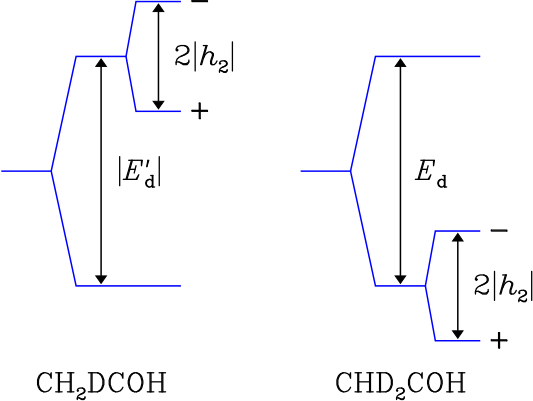}
\caption{\label{patt}The $J=0$ tunneling pattern of
\ce{CH2DCHO} and \ce{CHD2CHO} as retrieved with \citet{margules:09}. The tunneling parameter $h_2$ and the energy differences $E_d$ and $E_{d}'$ are defined in Section~\ref{theo}. The tunneling sublevels for \ce{CHD2CHO} are also labeled with the quantum number $\upsilon$ such that $\upsilon$ = 0 and 1 correspond respectively to the + and - tunneling sublevels and $\upsilon$ = 2 to the In conformation level.}\end{figure}

When tunneling effects are small, the In configuration displays asymmetric-top rotational energies shifted by $+E_d$.  For the $+$ and $-$ sublevels arising from the Out configurations, Eq.~(21) of \citet{margules:09} shows that $\pm h_2$ should be added to the asymmetric-top rotational energies, where the upper (lower) sign is for the $+$ ($-$) sublevel. As $h_2$ is negative \citep{hougen:85, coudert:88}, the $+$ sublevel is below the $-$ sublevel. The resulting tunneling pattern for $J=0$ is drawn in Fig.~\ref{patt} where it is compared to that of the
monodeuterated species \ce{CH2DCHO}. In agreement with the energy level diagram for \ce{CHD2CHO} in this figure, the vibrational
quantum number $\upsilon$, with 0 $\leq$ $\upsilon$ $\leq$ 2, is introduced. $\upsilon$ = 0 and 1 refer to rotational levels arising from the + and -- tunneling sublevels, respectively, and $\upsilon$ = 2 those arising from the In configuration. The results presented for \ce{CH2DCHO} by \citet{coudert:19} concerning selection rules, distortion terms to the tunneling matrix elements, and the assignment of the levels arising from numerical diagonalisation of the Hamiltonian matrix also apply for \ce{CHD2CHO} and the reader are referred to that paper for further information.

%\begin{table}
%\caption{\label{rotdep}Calculated molecular parameters}
%\centering
%\begin{tabular}{lclc}\hline\hline
%Parameter & Value & Parameter & Value \\ \hline
%$\chi_2$   & 241.9 & $\chi_3$   & \spcn{0}241.3 \\
%
%$\theta_2$ & \spcn{2}4.7 & $\theta_3$ & \spcn{2}3.7 \\
%
%$\phi_2$   & \spcn{1}61.9 & $\phi_3$   & \spcn{1}68.9 \\ \hline
%
%$A^{{\rm In}}$  & 1.599 & $A^{{\rm Out}}$ & 1.506 \\
%$B^{{\rm In}}$  & 0.293 & $B^{{\rm Out}}$ & 0.306 \\
%$C^{{\rm In}}$  & 0.272 & $C^{{\rm Out}}$ & 0.273 \\ \hline
%
%$\mu_x^{{\rm In}}$ & 1.043 & $\mu_x^{{\rm Out}}$ & 1.136 \\
%$\mu_y^{{\rm In}}$ & -     & $\mu_y^{{\rm Out}}$ & 0.111 \\
%$\mu_z^{{\rm In}}$ & 2.544 & $\mu_z^{{\rm Out}}$ & 2.502 \\
%\hline
%\end{tabular}
%\tablefoot{Eulerian-type angles, in degrees, involved in
%the rotational dependence of the tunneling matrix elements,
%the rotational constants, in cm$^{-1}$, and the dipole
%moments components, in Debye, are listed for the In and
%Out configurations. For symmetry reason, the relation
%$\chi_2=\phi_2 +\pi$ is fulfilled and $\mu_y^{{\rm In}}$
%is zero.  Superscripted In and Out labels identify the
%rotational constants and dipole moment components.}
%\end{table}

\begin{table}
\caption{\label{rotdep}Calculated molecular parameters}
\centering
\begin{tabular}{lclc}\hline\hline
Parameter & Value & Parameter & Value \\ \hline
$\chi_2$   & 241.9 & $\chi_3$   & \spcn{0}241.3 \\

$\theta_2$ & \spcn{2}4.7 & $\theta_3$ & \spcn{2}3.7 \\

$\phi_2$   & \spcn{1}61.9 & $\phi_3$   & \spcn{1}68.9 \\ \hline

$A^{{\rm In}}\times10^{-3}$  & 47.937 & $A^{{\rm Out}}\times10^{-3}$ & 45.149 \\
$B^{{\rm In}}\times10^{-3}$  & 8.784 & $B^{{\rm Out}}\times10^{-3}$ & 9.174 \\
$C^{{\rm In}}\times10^{-3}$  & 8.154 & $C^{{\rm Out}}\times10^{-3}$ & 8.184 \\ \hline

$\mu_x^{{\rm In}}$ & 1.043 & $\mu_x^{{\rm Out}}$ & 1.136 \\
$\mu_y^{{\rm In}}$ & -     & $\mu_y^{{\rm Out}}$ & 0.111 \\
$\mu_z^{{\rm In}}$ & 2.544 & $\mu_z^{{\rm Out}}$ & 2.502 \\
\hline
\end{tabular}
\tablefoot{Eulerian-type angles, in degrees, involved in
the rotational dependence of the tunneling matrix elements,
the rotational constants, in MHz, and the dipole
moments components, in Debye, are listed for the In and
Out configurations. For symmetry reason, the relation
$\chi_2=\phi_2 +\pi$ is fulfilled and $\mu_y^{{\rm In}}$
is zero.  Superscripted In and Out labels identify the
rotational constants and dipole moment components.}
\end{table}

\subsection{\label{anal_all}Line assignment and line analysis}
Starting from the results of \citet{turner:76}, parallel
$a$-type and perpendicular $b$-type transitions within the In configuration were assigned up to $J=20$ and $K_a=5$.  This first set of transitions was fitted with a Watson-type Hamiltonian. Parallel $a$-type and perpendicular $b$- and $c$-type transitions within and between the $+$ and $-$ sublevels of the Out configurations were afterwards assigned up to $J=27$ and $K_a=16$, using the results of \citet{turner:81}. Fitting of this second set of transitions yielded rotational constants for the Out configurations, the magnitude of the tunneling splitting $h_2$, and the Eulerian-type angles $\theta_2$ and $\phi_2$.  No unaccountably large residuals, which could have been attributed to couplings between the In and Out configurations, were found. As a result, unlike in the monodeuterated species \ce{CH2DCHO}, the value of $E_d$ and of the parameters describing the $1\rightarrow3$ tunneling parameter could not be retrieved.  Both sets of transitions were then fitted and new transitions were predicted and searched for. For the In configuration, it was possible to assign $a$- and $b$-type transitions up to $J=26$ and $K_a=17$.  For the Out configurations, $a$-, $b$-, and $c$-type transitions were assigned up to $J=27$ and $K_a=14$.  Table~\ref{countrans} lists the number of assigned transitions for each configuration counting forbidden even $\Delta K_a$ and $\Delta K_c$ transitions \citep{turner:81} of the Out configurations as $a$-type transitions.\ 
\begin{table}[H]
\caption{\label{countrans}Assigned transitions}
\centering
\begin{tabular}{lcc c cccc}\hline\hline
 & \multicolumn{2}{c}{In} && \multicolumn{3}{c}{Out} & \mbox{} \\ \cline{2-3} \cline{5-7}
References & \makebox[2.0em][c]{$a$-type} & \makebox[2.0em][c]{$b$-type} & &
\makebox[2.0em][c]{$a$-type} & \makebox[2.0em][c]{$b$-type} & \makebox[2.0em][c]{$c$-type} & All \\ \hline
1 & \spcn{1}17 & \spcn{1}30 && - & - & - & \spcn{1}47 \\
2 & - & - && \spcn{1}24 & \spcn{1}28 & \spcn{1}12 & \spcn{1}64 \\
{\em This work} & \spcn{0}176 & \spcn{1}36 && \spcn{0}445 & \spcn{1}75 & \spcn{2}2 & \spcn{0}742 \\ \hline
All & \spcn{0}193 & \spcn{1}66 && \spcn{0}469 & \spcn{0}103 & \spcn{1}14 & \spcn{0}853 \\ \hline
\end{tabular}
\tablebib{(1)~\citet{turner:76}; (2)~\citet{turner:81}}
\tablefoot{The number of assigned $a$-, $b$-, and $c$-type transitions
for each configuration in the two previous
investigations \citep{turner:76, turner:81} and in this
work. $c$-type transitions within the In configuration are
not allowed. No transitions were assigned between the In and Out configurations.}
\end{table}
In the final analysis, experimental frequencies were introduced in a least-squares fit procedure where they were given a weight equal to the inverse of their experimental uncertainty squared. Unresolved $K$-type doublets were treated as in \citet{margules:09}. The rotational Watson-type Hamiltonians used for the In and Out configurations were written using Watson's $A$-set of
distortion parameters \citep{watson:a, watson:b, watson:c}. The root mean square value of the observed minus calculated frequency is 81~kHz for transitions within the In configuration, 88~kHz for transitions within the Out configurations, and 83~kHz for all transitions.  The unitless standard deviation of this final analysis is 1.7. With the selected set of spectroscopic parameters, most line frequencies are reproduced within their experimental uncertainty of 50 kHz. $a$-type lines characterised by large $J$- and $K_a$-value tend to display residuals larger than this value and this may be due to the unaccounted for effects of the 1 $\rightarrow$ 3 tunneling motion. \refstepcounter{table}\label{table_omc} For the whole dataset, assignments, observed and calculated frequencies, and residuals are listed in Table~\ref{table_omc}, available at the Centre de Données astronomiques de Strasbourg (CDS)\footnote{\url{https://cdsweb.u-strasbg.fr/index-fr.gml}}.
This table displays 13 columns: Columns~1 to 4 (5 to 8) give the assignment of the upper (lower) level in terms of $J, K_{a}, K_{c}$ rotational quantum numbers and the vibrational state number $\upsilon$;  column~9 is the observed frequency in MHz; column~10 its uncertainty in kHz; column~11 is the observed minus calculated residual in kHz; column~12 is blank for a single line and d for a line belonging to an unresolved $K$-type doublet; and column~13 gives the reference from which the transition was taken.  Table~\ref{table_param_all} lists the parameters determined in the analysis. This table displays 2 columns.  Column~1 gives the parameter name; column~2 its value and uncertainty. For the rotational constants, the calculated values in Table~\ref{rotdep} are within 300 MHz from the experimental values in Table~\ref{table_param_all}.  For the Eulerian-type angles describing the rotational dependence of the tunneling matrix elements, the discrepancies are 0.2 and 1\dgr\ for, respectively, $\theta_2$ and $\phi_2$.\ 

\begin{centering}
\begin{threeparttable}
\footnotesize
\caption{\label{table_param_all}Spectroscopic parameters}
\begin{tabular}{llr@{.}l@{\hspace*{1.5\tabcolsep}}llr@{.}l}
\multicolumn{2}{l}{Parameter\tnote{a}} & \multicolumn{2}{c}{Value} & \multicolumn{2}{l}{Parameter\tnote{a}} & \multicolumn{2}{c}{Value} \\ \hline
$\rule{0.0em}{3.0ex}
\theta_2$ & & 4&864(12) &
$H_{K\!K\!J}$ & $\times 10^{4}$ & \minnumb0&514(21) \\
$\phi_2$ & & 60&828~7(36) &
$H_{J\!K\!K}$ & $\times 10^{5}$ & \minnumb0&112(81) \\
$\theta_{2j}$ & $\times 10^{3}$ & 0&293(34) &
$H_{J}$ & $\times 10^{8}$ & 0&26(25) \\
$\phi_{2j}$ & $\times 10^{3}$ & 0&187(40) &
$h_K$ & $\times 10^{3}$ & \minnumb0&434(63) \\
$h_2$ & & \minnumb91&824(22) &
$h_{K\!J}$ & $\times 10^{5}$ & \minnumb0&396(39) \\
$h_{2k}$ & & \minnumb0&116~7(23) &
$h_J$ & $\times 10^{9}$ & 0&7(13) \\
$h_{2j}$ & $\times 10^{2}$ & 0&949(81) &
\multicolumn{4}{c}{\mbox{}} \\
$f_2$ & $\times 10$ & 0&180~7(72) &
$A^{{\rm Out}}$ & $\times10^{-3}$& 45&141~639~0(39) \\
$s_{2xz}$ & $\times 10$ & \minnumb0&194(11) &
$B^{{\rm Out}}$ & $\times10^{-3}$& 9&176~200~44(90) \\
$h_{2kk}$ & $\times 10^{3}$ & \minnumb0&298(20) &
$C^{{\rm Out}}$ & $\times10^{-3}$& 8&187~146~15(81) \\
$h_{2kj}$ & $\times 10^{3}$ & 0&130~7(81) &
$\Delta_{K}$ & & 0&353~84(15) \\
$h_{2jj}$ & $\times 10^{5}$ & \minnumb0&277(81) &
$\Delta_{K\!J}$ & $\times 10$ & 0&168~73(14) \\
\multicolumn{4}{c}{\mbox{}} &
$\Delta_{J}$ & $\times 10^{2}$ & 0&666~13(20) \\
$A^{{\rm In}}$ & $\times10^{-3}$& 47&940~411~5(57) &
$\delta_K$ & $\times 10$ & \minnumb0&902~4(17) \\
$B^{{\rm In}}$ & $\times10^{-3}$ & 8&778~690~6(14) &
$\delta_J$ & $\times 10^{3}$ & 0&853~2(12) \\
$C^{{\rm In}}$ & $\times10^{-3}$& 8&175~064~5(13) &
$H_{K\!K\!J}$ & $\times 10^{4}$ & \minnumb0&238~0(90) \\
$\Delta_{K}$ & & 0&292~10(24) &
$H_{J\!K\!K}$ & $\times 10^{6}$ & \minnumb0&31(29) \\
$\Delta_{K\!J}$ & $\times10$& 0&778~46(42) &
$H_{J}$ & $\times 10^{8}$ & 0&70(22) \\
$\Delta_{J}$ & $\times10^{2}$& 0&517~80(26) &
$h_K$ & $\times 10^{5}$ & \minnumb0&3~174(13) \\
$\delta_K$ & & \minnumb0&353~56(51) &
$h_{K\!J}$ & $\times 10^{5}$ & \minnumb0&113~(21) \\
$\delta_J$ & $\times 10^{3}$ & 0&592~2(14) &
$h_J$ & $\times 10^{8}$ & 0&16(12) \\\hline
\end{tabular}
\begin{tablenotes}
\item[a]Parameters are in MHz except for the angles
$\theta_2, \phi_2$ and their distortion constants which are in degrees. Uncertainties are given in parentheses in the same units as the last quoted digit.
\end{tablenotes}
\end{threeparttable}
\end{centering}

%\begin{table}
%\caption{\label{table_param}Spectroscopic parameters}
%\centering
%\begin{tabular}{llll}\hline\hline
%\makebox[2.0em][l]{Parameter} & \multicolumn{1}{c}{Value} &
%\makebox[2.0em][l]{Parameter} & \multicolumn{1}{c}{Value} \\ \hline
%$A^{{\rm In}}$  & 1.599~120~4(2)  &
%$A^{{\rm Out}}$ & 1.505~763~4(1)  \\
%
%$B^{{\rm In}}$  & 0.292~825~60(5) &
%$B^{{\rm Out}}$ & 0.306~085~07(3) \\
%
%$C^{{\rm In}}$  & 0.272~690~84(4) &
%$C^{{\rm Out}}$ & 0.273~093~78(3) \\
%
%$h_2 \times 10^{3}$ & \minnumb3.062~9(7) \\
%
%$\theta_2$          &          4.864(12)  &
%$\phi_2$            & \spcn{-1}60.828(3)    \\ \hline
%\end{tabular}
%\tablefoot{Lowest order parameters obtained in the
%line position analysis of Section~\ref{anal_all}.  Parameters are in cm$^{-1}$ except %for
%the angles $\theta_2$ and $\phi_2$
%which are in degrees. Uncertainties are given in parentheses in the same units as the %last quoted digit.}
%\end{table}

\begin{table}[H]
\caption{\label{Q_values}Partition functions ($Q_{{\rm rot}}$) of \ce{CHD2CHO}} 
\centering
\begin{tabular}{cccc}
\hline\hline
$T/\mbox{K}$  & $Q_{{\rm rot}}$ &
$T/\mbox{K}$  & $Q_{{\rm rot}}$ \\ \hline
\spcn{2}2.725 & \spcn{3}27.3   & \spcn{1}75 & \spcn{1}5186.2 \\

\spcn{2}5.000 & \spcn{3}66.9   & 150 &        15262.5 \\

\spcn{2}9.375 & \spcn{2}177.5  & 225 &        28438.5 \\

\spcn{1}18.75 & \spcn{2}550.6  & 300 &        44014.5 \\

\spcn{1}37.50 & \spcn{1}1714.4 \\
\hline
\end{tabular}
\tablefoot{The partition functions are given for each temperature $T$ in Kelvin.}\end{table}

\subsection{Spectroscopic catalogue}
The spectroscopic catalogue was built using the results of the previous sections.  The energy difference $E_d$ was set to the value computed in Section~\ref{theo}.
Transitions were calculated up to $J=28$ and
their line strength and line intensity were computed using the dipole moment components in Table~\ref{rotdep}.  The partition functions $Q_{{\rm rot}}$, listed in Table~\ref{Q_values}, were computed for several temperatures using degeneracy factors equal to $(2J+1)$.  A zero energy was taken for the Out configurations $0_{00},+$ level.  Lines were selected using an intensity cutoff depending on the line frequency (as commonly done in the Jet Propulsion Laboratory (JPL) database catalogue line files; \citealp{jpl_dtb}). Its value in $\mbox{nm}^2\cdot\mbox{MHz}$ units at 300\,K is
\begin{equation}
10^{\mbox{{\small LOGSTR0}}} + (F/300~000)^2\times
10^{\mbox{{\small LOGSTR1}}},
\label{CUTOFF}
\end{equation}
where $F$ is the frequency in MHz, and {\small LOGSTR0} and
{\small LOGSTR1} are two dimensionless constants both set to $-8$.\refstepcounter{table}\label{table_jpl_asym} The linelist, given in Table~\ref{table_jpl_asym}, is available at the CDS and is formatted in the same way as the catalogue line files of the JPL database \citep{jpl_dtb}. Columns 1 to 3 contain, respectively, the line frequency (FREQ) in MHz, the error (ERR) in MHz, and the base 10 logarithm of the line intensity (LGINT) in $\mbox{nm}^2\cdot\mbox{MHz}$ units at 300\,K. Columns 4 to 6 give the degrees of freedom of the rotational partition function (DR), the lower state energy (ELO) in cm$^{-1}$, and the upper state degeneracy (GUP), respectively. Columns 7 and 8 contain the species tag (TAG) and format number (QNFMT), respectively. Finally, columns 9 to 12 (13 to 16) give the assignment of the upper (lower) level in terms of $J, K_{a}, K_{c}$, and the vibrational state number $\upsilon$.  A minimum value of 10~kHz was selected for the calculated error (ERR).  For observed unblended microwave lines, the line frequency (FREQ) and the error (ERR) were replaced by their experimental values. This is then indicated by a negative species tag. The catalogue will be available on CDMS \footnote{\url{https://cdms.astro.uni-koeln.de}} \citep{endres:16}.

\section{Astrophysical observations} \label{obs}

Based on the new spectroscopic measurements, we searched for \dda \, towards the B component of the protostellar system IRAS~16293--2422 in data from the ALMA Protostellar Interferometric Line Survey, PILS \citep{jorgensen:16}. PILS represents an unbiased molecular line survey of IRAS 16293--2422 carried out during ALMA's Cycle 2 (project id: 2013.1.00278.S, PI: J. K. J{\o}rgensen) covering one of the prominent atmospheric windows in ALMA's Band 7 between 329.1 and 362.9~GHz with a spectral resolution of $\approx$0.2~km~s$^{-1}$ and angular resolution of $\approx$0.5$''$ (70~au at the distance of IRAS~16293--2422). The high sensitivity of the PILS data and relatively narrow lines towards one component of IRAS~16293--2422 have enabled the detections of a number of species for the first time in the ISM (e.g. CH$_3$Cl by \cite{fayolle:17} and HONO by \cite{coutens:19}) as well as making it possible to systematically survey the content of deuterated isotopologues of complex organic molecules \citep{jorgensen:18}. The latter also includes the detection of doubly-deuterated organics including CHD$_2$CN \citep{calcutt:18}, CHD$_2$OCHO \citep{manigand:19} and CHD$_2$OCH$_3$ \citep{richard:21} as well as new and better constraints on the column densities of doubly- and triply-deuterated methanol \citep{drozdovskaya:22, ilyushin:22}. For details about the data and their reduction we refer to \cite{jorgensen:16}.

\begin{figure*}[ht]
\centering
\includegraphics[width=20cm]{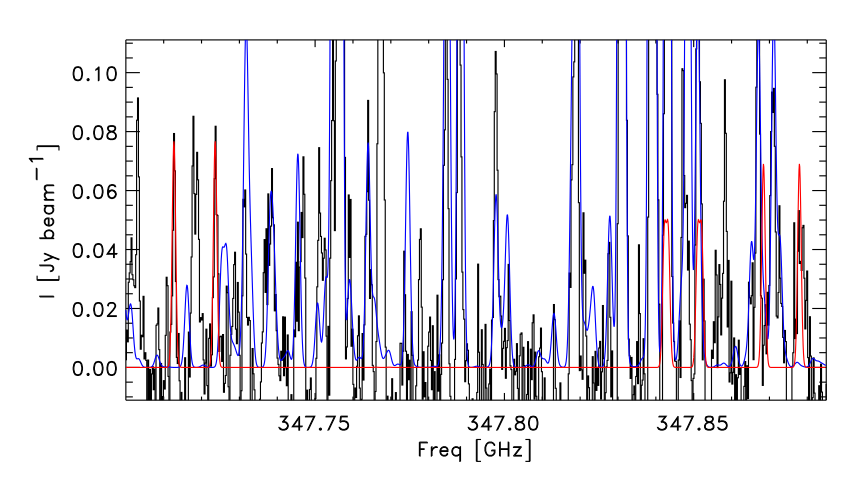}
\caption{Example of the CHD$_2$CHO fits in a selected frequency range. The synthetic spectra fitted to CHD$_2$CHO is shown in red and all other species identified in PILS with blue.}
\label{specgen}
\end{figure*}

For our search we analysed the position offset by one beam (0.5$''$) from the B component of IRAS~16293--2422, where line and continuum opacity is limited. This position was also the one studied in the above-mentioned papers from PILS \citep{jorgensen:16, jorgensen:18}. We adopted a similar approach to previous works by fitting synthetic spectra for \dda \, calculated under the assumption that its excitation is characterised by local thermodynamical equilibrium (LTE), which is reasonable at the densities of the warm gas where these species are present \citep{jorgensen:16}. The free parameters in the fits are the column density of the molecule $N$, and its rotational temperature, $T_{\rm rot}$. For the line width and velocity offset relative to the local standard of rest we adopt values of 1~km~s$^{-1}$ (FWHM) and of 2.6~km~s$^{-1}$ respectively, which match the spectra well at this position. For $T_{\rm rot}$ we assumed a temperature of 125~K similar to that of the non-deuterated and singly-deuterated isotopologues of acetaldehyde. An example of the fit to a selected frequency range is shown in Fig.~\ref{specgen} while the fits to the lines predicted to be brighter than 40~mJy~beam$^{-1}$~km~s$^{-1}$ (68 transitions; the RMS noise in the spectra is about 4--5~mJy~beam$^{-1}$ per 1 km~s$^{-1}$) over the entire frequency range are shown in Fig.~\ref{spec}-\ref{spec3} [in Appendix \ref{add}]. Several clean and unblended transitions are seen providing a good constraint on the CHD$_2$CHO column density of 1.3$\times 10^{15}$~cm$^{-2}$ with an uncertainty of 10-20\% (for the discussion on the uncertainty derivation we refer to \cite{jorgensen:18}). The few lines that are either under- or over-produced with the synthetic spectra are due to blends with brighter lines of more prominent species (e.g the two lines seen at 330.71~GHz with an upper energy level of 202~K that are blended with glycolaldehyde) or absorption due to optically thick emission (e.g. the transition at 347.86~GHz falling close to a transition of formic acid).

The derived column density can be compared to that of the singly deuterated variant, CH$_2$DCHO, of $6.2\times10^{15}$ cm$^{-2}$ \citep{manigand:20}. The ratio between the singly and doubly deuterated variants of 20\% is very close to those for methyl formate (CHD$_2$OCHO/CH$_2$DOCHO) of 22\% \citep{manigand:19}, dimethylether (CH$_3$OCHD$_2$/CH$_2$OCH$_2$D) of 15-20\% \citep{richard:21} and methanol (CHD$_{2}$OH/CH$_{2}$DOH) of 25\% \citep{drozdovskaya:22} -- in all cases significantly above the ratios for the singly-deuterated to non-deuterated isotopologues \citep{jorgensen:18} by factors of 4--5.

\section{Discussion of astronomical observational results}\label{discussion}

The similar doubly to singly deuterated column density ratios for acetaldehyde, methyl formate, dimethylether and methanol presented in Section \ref{obs} suggest doubly deuterated acetaldehyde shares a common origin and was formed in an environment resembling the physical conditions with which doubly deuterated methyl formate, dimethylether and methanol were formed. The pre-stellar core phase is a good candidate due to the low temperatures that promote deuterium fractionation through the enhancement of the H$_{2}$D$^{+}$/ H$_{3}^{+}$ ratio as well as the larger atomic D/H ratio in the gas phase, which promotes deuteration of surface species.\

Mechanisms for acetaldehyde main isotopologue formation have been suggested both for the gas phase as well as for the surface of grains. For the gas phase \cite{vazart:20} concluded that C$_{2}$H$_{5}$ + O($^{3}$P) and CH$_{3}$CH$_{2}$OH + OH / CH$_{3}$CHOH + O($^{3}$P) are potentially efficient gas-phase formation routes. \cite{fedoseev:22} studied grain surface reactions and proposed CH$_{2}$CO + 2H as a plausible way in forming CH$_{3}$CHO. Contrarily to deuterated methanol, whose formation pathway has been constrained \citep{drozdovskaya:21}, the formation of deuterated acetaldehyde is still not clear. In the same line as doubly deuterated methanol, doubly deuterated acetaldehyde could be formed from a doubly deuterated reactant. The acetaldehyde D$_{2}$/D ratio found in this work and the D/H ratio from \citealp{manigand:20} combined with a gas-grain chemical model can potentially shed light on the formation mechanism of singly- and multi-deuterated acetaldehyde.\ 

Contrary to deuterated methanol, which has been observed in a variety of sources, deuterated acetaldehyde counts less detections. Acetaldehyde is less abundant than methanol in pre-stellar cores with a [CH$_{3}$CHO]/[CH$_{3}$OH] ratios betweewn 0.02 and 0.26 \citep{scibelli:20}. We estimate the line brightness of the most intense lines of the singly and doubly deuterated acetaldehyde in the 3 mm band towards pre-stellar cores to be 4.3 mK and 0.5 mK, respectively. 
We base our estimate on the average column density of \ce{CH3CHO} observed towards starless cores in \citealp{scibelli:20} (2$\times10^{12}$ cm$^{-2}$), and assuming the D/H and D$_{2}$/H ratio observed towards IRAS16293.\ 

Based on our predictions it will be unlikely to detect the doubly deuterated acetaldehyde towards pre-stellar cores, and this is possibly true also for other COMs. Nevertheless, one can use the diagnostic power of deuteration and  derive the information on inheritance from pre-stellar cores by using state-of-the-art chemical models.

\section{Conclusions} \label{concl}

The rotation-torsion spectrum of doubly deuterated acetaldehyde (\dda) \ was experimentally and theoretically studied. Due to the tunneling of the CHD$_{2}$ methyl group, the ground state is split in three torsional sublevels. Transitions were measured in the millimetre and submillimetre range (82.5 -- 450 GHz), as described in Section \ref{exp}. These, alongside previously measured ones, were fitted using the internal axis method (IAM). A total of 853 transitions were fitted with a weighted root mean square deviation of 1.7. The resulting spectroscopic parameters computed with this fit can be found in Table \ref{table_param_all}. We build a spectroscopic catalogue for astrophysical purposes from the results of the analysis, which we publish in CDMS.\ 

We present the first detection of \ce{CHD2CHO} in the interstellar medium through comparisons to observations of the B component of the protostellar system IRAS 16293-2422 from the ALMA PILS program. This doubly deuterated variant is enhanced compared to its singly- and non-deuterated counterparts at the same level as for other complex organics seen towards this source. Further comparison to chemical models may shed further light on the acetaldehyde formation during the earliest stages of star formation.\

\section*{Acknowledgements.} J.F.A., S.S., V.L., C.P.E. and P.C. gratefully acknowledge the support of the Max Planck Society. The research of J.K.J. is supported by the Independent Research Fund Denmark (grant No. 0135-00123B). This paper makes use of the following ALMA data: ADS/JAO.ALMA$\#$2013.0.00278.S. ALMA is a partnership of ESO (representing its member states), NSF (USA) and NINS (Japan), together with NRC (Canada), MOST and ASIAA (Taiwan), and KASI (Republic of Korea), in cooperation with the Republic of Chile. The Joint ALMA Observatory is operated by ESO, AUI/NRAO and NAOJ. We also thank the anonymous referee for their insightful comments.

\bibliographystyle{aa}
\bibliography{vanderwaals}

\onecolumn
\appendix

\section{Additional material}\label{add}
\begin{figure}[ht]
\includegraphics[width=\columnwidth]{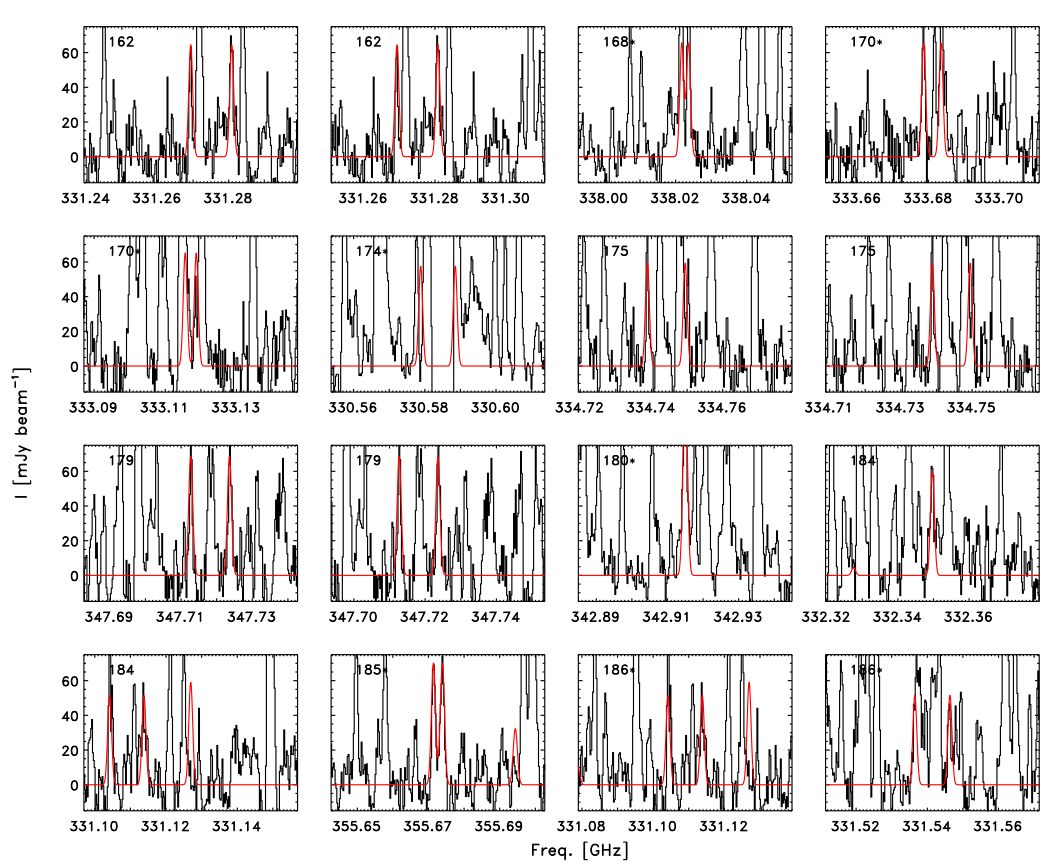}
\caption{The 68 transitions of CHD$_2$CHO predicted to be the brightest assuming a rotational temperature of 125~K. The red line indicates the predicted line intensities obtained by fitting to the lines with the synthetic spectra thereby constraining the column density. In each panel, the numbers on the upper left corner indicate the excitation temperature $T_{ex}$ of the fitted transitions. An asterisk next to this number indicates situations where two lines from Table \ref{detectedtrans} with similar values for $E_u$ fall within 10 MHz of each other and are shown together in one panel.}
\label{spec}
\end{figure}

In this section we present Table \ref{detectedtrans} listing the transitions of \ce{CHD2CHO} detected in the PILS ALMA Band 7 frequency range. Moreover, we present the fits of these transitions assuming a rotational temperature of 125 K over the PILS ALMA Band 7 spectrum in Figures \ref{spec} to \ref{spec3}.

\begin{table}[ht]
\caption{\label{detectedtrans} \ce{CHD2CHO} transitions with predicted intensities above 40 mJy~beam$^{-1}$~km~s$^{-1}$ shown in Figures \ref{spec}--\ref{spec3} with the fitted column density and excitation temperature described in Sect. \ref{obs}. The frequencies appear in MHz and the upper energies in K.}
\centering
\begin{tabular}{lll|lll}
\hline\hline
Transition & Frequency & E$_{up}$ & Transition & Frequency & E$_{up}$ \\
$J^{'}_{K_{a}^{'},K_{c}^{'},v^{'}} \rightarrow J_{K_{a},K_{c},v}$ & MHz & K & $J^{'}_{K_{a}^{'},K_{c}^{'},v^{'}} \rightarrow J_{K_{a},K_{c},v}$ & MHz & K \\
\hline

19$_{3,17,0}$ $\rightarrow$ 18$_{3,16,0}$  &  330578.85    &   174 & 20$_{1,19,1}$ $\rightarrow$ 19$_{1,18,1}$ &  347712.71    &   179 \\
19$_{3,17,1}$ $\rightarrow$ 18$_{3,16,1}$  &  330588.53    &   174 & 20$_{1,19,0}$ $\rightarrow$ 19$_{1,18,0}$ &  347723.60    &   179  \\
19$_{5,14,0}$ $\rightarrow$ 18$_{5,13,0}$  &  330710.33    &   202 & 20$_{6,14,1}$ $\rightarrow$ 19$_{6,13,1}$ &  347843.28 & 238 \\
19$_{5,15,0}$ $\rightarrow$ 18$_{5,14,0}$  &  330712.14    &   202 & 20$_{6,15,0}$ $\rightarrow$ 19$_{6,14,0}$ &  347851.04 & 238 \\
19$_{5,14,1}$ $\rightarrow$ 18$_{5,13,1}$  &  330721.99    &   202 & 20$_{3,18,0}$ $\rightarrow$ 19$_{3,17,0}$ &  347868.44 &  191 \\
19$_{5,15,1}$ $\rightarrow$ 18$_{5,14,1}$  &  330723.84    &   202 & 20$_{3,18,1}$ $\rightarrow$ 19$_{3,17,1}$ &  347877.93 &  191 \\
19$_{4,16,0}$ $\rightarrow$ 18$_{4,15,0}$  &  331104.17    &   186 & 20$_{5,16,0}$ $\rightarrow$ 19$_{5,15,0}$ &  348199.22    &   218 \\
19$_{4,16,1}$ $\rightarrow$ 18$_{4,15,1}$  &  331113.81    &   186 &  20$_{5,15,0}$ $\rightarrow$ 19$_{5,14,0}$$^*$ &  348211.24 &  218 \\
20$_{1,20,2}$ $\rightarrow$ 19$_{1,19,2}$  &  331126.94    &   184 & 20$_{5,16,1}$ $\rightarrow$ 19$_{5,15,1}$$^*$ &  348212.14 &  218 \\
19$_{1,18,1}$ $\rightarrow$ 18$_{1,17,1}$  &  331269.34    &   162 & 20$_{5,15,1}$ $\rightarrow$ 19$_{5,14,1}$ &  348224.15    &   218 \\ 
19$_{1,18,0}$ $\rightarrow$ 18$_{1,17,0}$  &  331280.85    &   162 & 21$_{0,21,2}$ $\rightarrow$ 20$_{0,20,2}$ &  348610.15    &   201 \\ 
19$_{4,15,0}$ $\rightarrow$ 18$_{4,14,0}$  &  331536.50    &   186 & 20$_{4,17,0}$ $\rightarrow$ 19$_{4,16,0}$ &  348616.55 &  203 \\  
19$_{4,15,1}$ $\rightarrow$ 18$_{4,14,1}$  &  331546.36    &   186 & 20$_{4,17,1}$ $\rightarrow$ 19$_{4,16,1}$ &  348626.53 &  203 \\
20$_{0,20,2}$ $\rightarrow$ 19$_{0,19,2}$  &  332349.70    &   184 & 20$_{4,16,0}$ $\rightarrow$ 19$_{4,15,0}$ &  349229.71    &   203 \\
20$_{1,20,1}$ $\rightarrow$ 19$_{1,19,1}$  &  333115.59    &   170 & 20$_{4,16,1}$ $\rightarrow$ 19$_{4,15,1}$ &  349239.90    &   203 \\ 
20$_{1,20,0}$ $\rightarrow$ 19$_{1,19,0}$  &  333118.70    &   170 & 21$_{1,21,1}$ $\rightarrow$ 20$_{1,20,1}$ &  349513.62 & 187  \\
20$_{0,20,1}$ $\rightarrow$ 19$_{0,19,1}$  &  333678.65    &   170 & 21$_{1,21,0}$ $\rightarrow$ 20$_{1,20,0}$ &  349516.59 & 187  \\
20$_{0,20,0}$ $\rightarrow$ 19$_{0,19,0}$  &  333683.53    &   170 & 21$_{0,21,1}$ $\rightarrow$ 20$_{0,20,1}$ &  349962.17 & 187  \\
19$_{2,17,1}$ $\rightarrow$ 18$_{2,16,1}$  &  338021.76    &   168 & 21$_{0,21,0}$ $\rightarrow$ 20$_{0,20,0}$ &  349966.60 & 187 \\
19$_{2,17,0}$ $\rightarrow$ 18$_{2,16,0}$  &  338023.63    &   168 & 20$_{3,17,0}$ $\rightarrow$ 19$_{3,16,0}$ &  352967.92  & 192 \\
20$_{5,16,2}$ $\rightarrow$ 19$_{5,15,2}$  &  339203.56    &   234 & 20$_{3,17,1}$ $\rightarrow$ 19$_{3,16,1}$ &  352977.86  & 192 \\
20$_{5,15,2}$ $\rightarrow$ 19$_{5,14,2}$  &  339205.81    &   234 & 21$_{2,20,2}$ $\rightarrow$ 20$_{2,19,2}$ &  354047.62  &   210 \\
20$_{2,19,1}$ $\rightarrow$ 19$_{2,18,1}$$^*$  &  342914.69    &   180 & 20$_{2,18,1}$ $\rightarrow$ 19$_{2,17,1}$ &  355671.41 & 185 \\
20$_{2,19,0}$ $\rightarrow$ 19$_{2,18,0}$$^*$  &  342915.25    &   180 & 20$_{2,18,0}$ $\rightarrow$ 19$_{2,17,0}$ &  355673.91 & 185 \\
19$_{3,16,0}$ $\rightarrow$ 18$_{3,15,0}$  &  334738.86    &   175 &  21$_{5,17,2}$ $\rightarrow$ 20$_{5,16,2}$ &  356182.55 & 251 \\
19$_{3,16,1}$ $\rightarrow$ 18$_{3,15,1}$  &  334749.47    &   175 &  21$_{5,16,2}$ $\rightarrow$ 20$_{5,15,2}$ &  356186.06 & 251 \\  
20$_{2,19,2}$ $\rightarrow$ 19$_{2,18,2}$  &  337358.80    &   193 &  21$_{3,19,2}$ $\rightarrow$ 20$_{3,18,2}$ &  356391.35    &   220 \\
20$_{4,17,2}$ $\rightarrow$ 19$_{4,16,2}$  &  339429.44    &   217 &  21$_{4,18,2}$ $\rightarrow$ 20$_{4,17,2}$ &  356433.69    &   234 \\ 
20$_{3,18,2}$ $\rightarrow$ 19$_{3,17,2}$  &  339437.37    &   203 &  21$_{4,17,2}$ $\rightarrow$ 20$_{4,16,2}$ &  356537.97    &   234 \\ 
20$_{4,16,2}$ $\rightarrow$ 19$_{4,15,2}$  &  339503.77    &   217 &  21$_{3,18,2}$ $\rightarrow$ 20$_{3,17,2}$ &  357931.72    &   221 \\ 
20$_{3,17,2}$ $\rightarrow$ 19$_{3,16,2}$  &  340666.65    &   204 &  21$_{1,20,2}$ $\rightarrow$ 20$_{1,19,2}$ &  358652.34    &   208 \\ 
20$_{1,19,2}$ $\rightarrow$ 19$_{1,18,2}$  &  341958.03    &   190 & 21$_{2,20,1}$ $\rightarrow$ 20$_{2,19,1}$$^*$ &  359639.51 & 198 \\ 
20$_{2,18,2}$ $\rightarrow$ 19$_{2,17,2}$  &  343844.42    &   195 & 21$_{2,20,0}$ $\rightarrow$ 20$_{2,19,0}$$^*$ &  359640.38 & 198 \\ 
21$_{1,21,2}$ $\rightarrow$ 20$_{1,20,2}$  &  347534.63    &   201 &  21$_{2,19,2}$ $\rightarrow$ 20$_{2,18,2}$ &  361170.20    &   213 \\
\hline
\end{tabular}

\tablefoot{$^*$ Transitions separated by less than 1 km~s$^{-1}$ (the FWHM of lines towards IRAS16293B) that consequently appear blended in Fig. \ref{spec}--\ref{spec3}.}

\end{table}

\begin{figure}[h]
\includegraphics[width=\columnwidth]{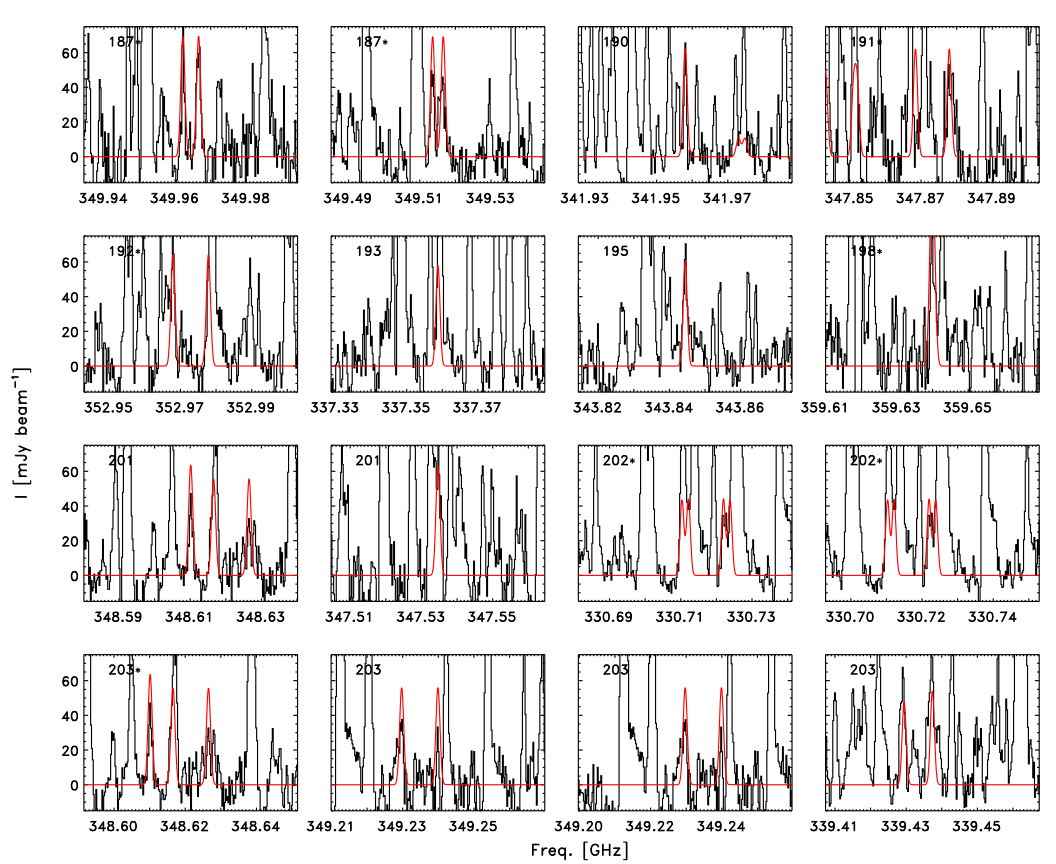}
\caption{Same as Figure \ref{spec}.}
\end{figure}

\begin{figure}[h]
\includegraphics[width=\columnwidth]{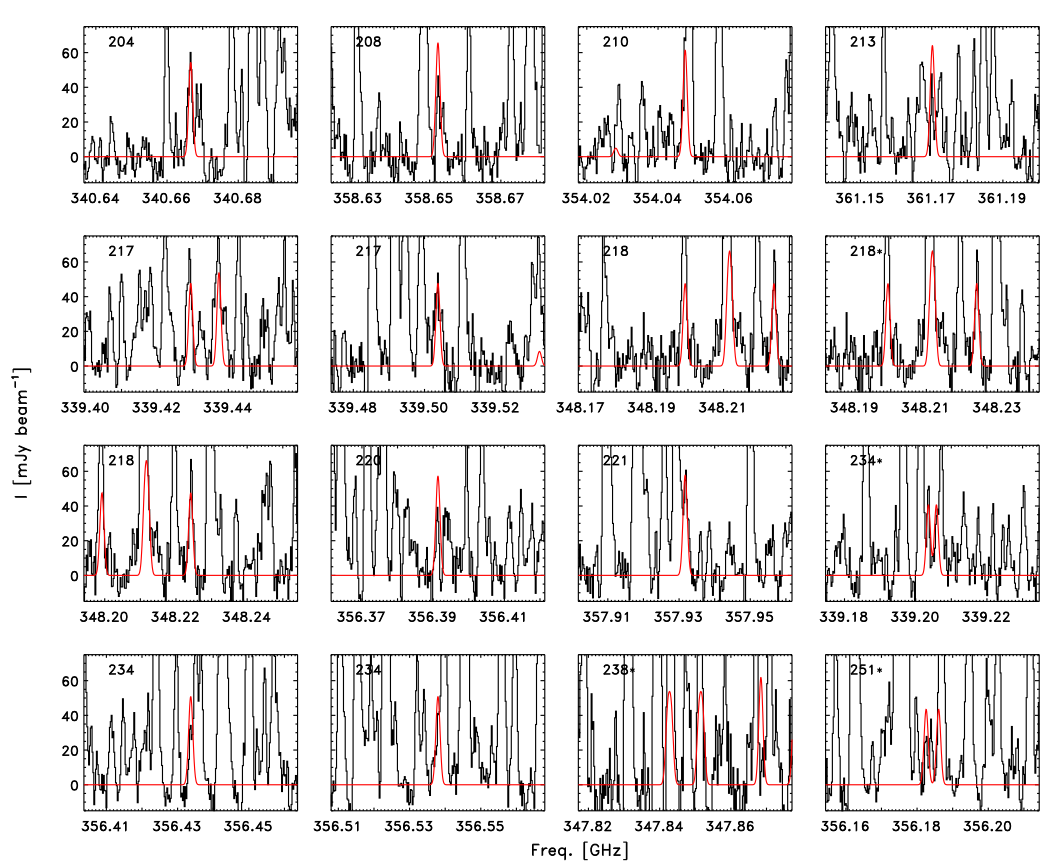}
\caption{Same as Figure \ref{spec}.}
\label{spec3}
\end{figure}

\end{document}